\documentclass[12pt,preprint]{aastex}

\newcommand{\swuma}{SW UMa}
\newcommand{\kms}{km s$^{-1}$}
\newcommand{\ovi}{\ion{O}{6}}
\newcommand{\cgs}{erg cm$^{-2}$ s$^{-1}$}
\newcommand{\lya}{Ly$\alpha$}
\newcommand{\ciii}{\ion{C}{3}}

\shorttitle{FUSE Observations of SW UMa}
\shortauthors{Povich et al.}

\begin{document}


\title{FUSE Observations of the Dwarf Nova \swuma \\ During Quiescence}


\author{M. S. Povich, J. C. Raymond, and A. Lobel}
\affil{Harvard-Smithsonian Center for Astrophysics, 60 Garden Street, Cambridge, MA 02138}

\and

\author{K. Menou\altaffilmark{1}}
\affil{Department of Astronomy, Columbia University, \\ 550 West 120th
Street, New York, NY 10027}

\altaffiltext{1}{On leave from Institut d'Astrophysique de Paris, 98bis
Boulevard Arago, 75014 Paris, France}


\begin{abstract}
We present spectroscopic observations of the short-period cataclysmic
variable 
SW Ursa Majoris, obtained
by the Far Ultraviolet Spectroscopic Explorer (FUSE)
satellite while the system was in quiescence. The data include the resonance
lines of \ovi\ at 1031.91 and 1037.61 \AA. These lines are present in
emission, and they exhibit both narrow ($\sim 150$ \kms) and broad ($\sim
2000$ \kms) components. The narrow \ovi\ emission lines exhibit
unusual double-peaked and redshifted profiles. We attribute the source of this
emission to a
cooling flow onto the surface of the white dwarf primary. The broad
\ovi\ emission most likely
originates in a thin, photoionized surface layer on the accretion disk. We
searched for emission from H$_{2}$ at 1050 and 1100 \AA, motivated by
the expectation that the bulk
of the quiescent accretion disk is in the form of cool, molecular
gas. 
If H$_{2}$ is present, then our limits on the fluxes of the H$_{2}$
lines are consistent with 
the presence of a surface layer of atomic H that shields
the interior of the disk. These results may indicate that 
accretion operates primarily in the surface layers of the disk in
\swuma.
We also investigate the far-UV continuum of \swuma\ and place an upper
limit of 15,000 K
on the effective temperature of the white dwarf.
\end{abstract}

\keywords{stars: binaries --- stars: dwarf novae ---
stars: individual (\objectname{SW UMa}) --- ultraviolet: stars}


\section{Introduction}\label{intro}

Cataclysmic variables (CVs) are binary star systems in which the primary,
a white dwarf (WD), 
accretes material from a Roche-lobe-filling secondary, usually
a late-type main-sequence star. In the absence of a strong magnetic
field from the WD, the stream flowing out from the secondary
forms an accretion disk around the primary. Such accretion disks serve
as reservoirs for the mass-transfer process;
thermal and viscous
instabilities cause the disk to switch to a high-temperature, mostly
ionized state from a low-temperature, mostly
neutral state. The transition from the high state back to the low
state is accomplished by rapid disposal of disk matter onto the WD. 
This disk-instability mechanism is
responsible for the periodic episodes of 
elevated rates of accretion onto the primary
that are the origin of the outbursts commonly observed in dwarf novae (DNe),
when the overall brightness of the system abruptly
increases by several magnitudes \citep{BW95,JC93,JL01}. 
During these eruptions, which can range from several days to a few
weeks in duration, 
the majority of the luminosity of the
system comes from the accretion disk and the associated 
hot spot (where the accretion stream from the secondary impacts the
outer edge of the disk) and boundary layer (where material from the
inner edge of the disk decelerates from its Keplerian velocity to meet
the more slowly rotating surface of the WD). Historically,
DNe have been classified on the basis of their lightcurves, and the
classifications are named after an observational archetype. The U Gem stars
are considered to be the typical DN systems. The lightcurves
of
SU UMa stars, an important subclass of U Gem stars,
exhibit periodic ``superoutbursts'' that are $\sim 0.7$ mag
brighter and $\sim 5$ times longer duration than usual. 

The presence of a strong magnetic field alters the geometry of the
mass-transfer process considerably. In AM Her systems, or ``polars,''
(see Cropper 1990 for a detailed review)
the field is
strong enough to prevent an accretion disk from forming altogether;
instead, the primary rotates synchronously with the orbital period
of the binary, and the accretion stream from the secondary is channeled along
the field lines directly down to the WD. AM Her systems 
do not exhibit dwarf nova eruptions, although they do alternate between
high and low accretion states. In systems where the magnetic field is
weaker, only the inner portion of the disk is disrupted, and material
may still collect in an outer disk or annulus 
before accreting onto the WD
along the field lines. Such systems are called ``intermediate polars'' (IPs).
The DQ Her stars are a subclass of the IPs, characterized by rapidly
rotating primaries.

SW UMa is a DN of the SU UMa subclass, and has been characterized as a
tremendous outburst amplitude dwarf nova \citep{RSHWM,TOAD}. 
It undergoes outbursts on a $\sim 460$-day cycle, jumping from
a minimum apparent magnitude $M_{V}=16.5$ during quiescence to a maximum
brightness of
$M_{V}=10.8$. 
It is an ultrashort-period binary with $P_{\rm{orb}} =
81.8$ min \citep{SIAU,SST86}. 
SST86 observed a 15.9-minute periodicity in the optical
photometry of \swuma\ in quiescence 
and also marginally detected this periodicity in 
soft X-rays (0.05--2 keV) 
observed by EXOSAT. These findings suggest that magnetic accretion is taking
place, and thus \swuma\ may be an IP. This idea is supported by
the fact that \swuma\ is an unusually strong X-ray emitter for
such a faint optical source.
ROSAT observations of the soft X-ray (0.1--2.5 keV) emission from \swuma\
in quiescence failed to detect the 15.9-minute period but could
not rule out its presence \citep{R94}. 
No 15.9-minute period has been detected in either high-speed optical
photometry \citep{RSHWM} or X-rays during superoutburst phases of \swuma, but this
could be due to the fact that, during outburst, the Alfv\'{e}n radius might
shrink to less than the WD radius \citep{SOH88}; in other words, the
greatly enhanced mass-transfer from the disk swamps the magnetic field
of the WD, and disk accretion then proceeds as in non-magnetic
systems.
Further searches for periodic behavior at the beat frequency
between the presumed 15.9-minute rotational period of the WD
and the 81.8-minute orbital period have failed to detect any
periodicity (P. Szkody, private communication, 2004). 
If indeed \swuma\ is an IP, it stands as one
of the two shortest-period examples known, the other being EX Hya.

In this paper we present observations of \swuma\ obtained by the
Far-Ultraviolet Spectroscopic Explorer (FUSE) while the system was in
quiescence. The original aim of these observations was to search for
emission from molecular hydrogen, which would serve as direct evidence that
the accretion disk cools sufficiently during quiescence to allow the
formation of molecules. This emission was not detected, but the
detection 
limits
leave open the possibility of layered accretion in the disk of
\swuma. 
We also looker for
a faint far-UV continuum from this source, 
which would indicate 
thermal (blackbody) emission from the WD. The upper limits on this
continuum can be used to constrain the post-outburst
cooling of \swuma. The only
obvious features in the FUSE spectra of \swuma\ are the \ovi\
resonance lines at 1031.91 \AA\ and 1037.61 \AA, which exhibit both
narrow and broad emission components. Broad emission of several blended 
transition of 
\ciii\ centered at 
1176 \AA\ is also detected. We examine the
unusual features of these lines and explore possible physical mechanisms
explaining their formation.

\section{Observations}\label{observe}

\subsection{FUSE Observations of \swuma}\label{swumaobs}

We obtained 10 exposures of \swuma\ with FUSE on JD 2452220 (2001
November 6, between 00:42 and 15:31 UT). These observations are
archived under FUSE guest investigator program b074. The FUSE instrument is
described in detail by \citet{M00} and \citet{S00}. The four FUSE
spectrographs cover the wavelength range of 905--1188 \AA, with
spectral channels denoted by combining the grating coating names (LiF
and SiC) with the telescopes (1 and 2) and detectors (a and b). Three
spectroscopic apertures are available (LWRS, MDRS, and HIRS),
providing monochromatic resolving powers of 20,000--24,000 ($\pm
2000$). \swuma\ was observed through the LWRS aperture, which has a
field of view that is 30$\arcsec$ square. 
The LWRS  
aperture center was placed at 129$^{\circ}$.178958 azimuth and 
+53$^{\circ}$.476972
declination
(provided in the exposure file header information), which corresponds to
SW UMa's coordinates of $\alpha_{2000}=08^{\rm h}$ $36^{\rm m}$
$42.95^{\rm s}$, $\delta_{2000}=+53^{\circ}$ $28\arcmin$ $37.1\arcsec$ 
to within an angular distance of 1.7$\arcsec$.      
The position angle of the y-axis of the LWRS aperture was $7.9^{\circ}$ 
East of North during the 10 FUSE science exposures of SW UMa. 
The photon events from \swuma\
were accumulated by the detectors in the time-tag (TTAG) mode. Because
of the low Earth orbit of FUSE (768 km), strong airglow emission
lines (primarily those of \ion{O}{1}, \ion{N}{1}, and the Lyman series
of \ion{H}{1}) 
due to scattered sunlight are present in the astrophysical
spectra observed with the large LWRS aperture during the daylight
portion of the orbit. 

We used the CalFUSE version 2.2.3 calibration software to recalibrate
the archival spectra for each exposure of \swuma. The 10 
exposures were then summed for each of the separate spectral
channels, yielding a single set of spectra representing 29,000 s
of exposure time.
The CalFUSE pipeline includes relative
wavelength calibration for the detector segments in each channel, and this
is believed to be accurate to within 2 pixels (0.012 \AA). The
absolute wavelength calibration is less reliable, so to 
correct for small pixel shifts between the subsequent
exposures we cross-correlated the LiF1a and LiF2b exposures over the
small wavelength range of 3 \AA\ centered on the \ovi\ 1031.9 \AA\
line. As a further check on 
the absolute wavelength scale in the vicinity of the
\ovi\ emission lines, we measured the centers of the 5 nearby \ion{O}{1}
airglow lines, which lie in the geocentric frame of rest. All 5 lines
were found to lie within 0.02 \AA\ (6 \kms) of their rest
wavelengths. We then transform the wavelength scale to heliocentric rest
by correcting for the -23 \kms\ shift introduced by Earth's
orbital motion in the direction of \swuma\ at the time of our observations.
The flux calibration steps include flat-fielding, background
subtraction, burst removal, and \linebreak 1-D spectrum extraction for absolute
flux conversion that corrects for optical (point-source) astigmatism
in the cross-dispersion direction.

\subsection{FUSE Sky-Background Observations Near \swuma}\label{bkgd}

Immediately prior to our observations of \swuma, FUSE obtained three
exposures with the
telescope apertures 
pointed at a blank patch of sky $\sim 1\arcmin$ to one side of
the target. These observations, part of the S405 ``sky-background''
program, total 9800 s of exposure time and provide a valuable
diagnostic of the ambient far-UV background spectrum in the immediate
vicinity of \swuma. We applied the same calibration steps to the
sky-background data as to the \swuma\ target data to ensure
consistency between both datasets.

\subsection{AAVSO lightcurves}\label{AAVSO}

The American Association of Variable Star
Observers (AAVSO) maintains an
extensive database of observations contributed by a gobal network of amateur
astronomers. These data provide invaluable lightcurve coverage on
countless variable stars. The AAVSO database
contains nearly 25,000 observations of \swuma, 
dating back to 1939. Using their
online lightcurve
generator,\footnote{\url{http://www.aavso.org/data/lcg}} it is 
straightforward to track the brightness of a source, and hence the 
outburst cycle of a DN like \swuma, over a
specified period of time. At the time of the FUSE observations of
\swuma\ 
(JD 2452220), the visual
magnitude of the system was $M_{V}\approx 17$. The previous and
subsequent eruptions, both marked by an increase in brightness to
$M_{V}\approx 10.5$, occurred 150 days before and 350 days after this
date. We are therefore confident that FUSE did indeed observe \swuma\
during its quiescent phase. 

\swuma\ was first observed in the UV 
by the International Ultraviolet Explorer (IUE) 
satellite 
on JD 2450057 (1995 5 December),
and while it is clear from the AAVSO lightcurves that the system was
in quiescence at that time,
it is uncertain when exactly the
previous DN eruption took place. The light curves
show that the system was in outburst on JD 2449200, more than 800 days
prior to the IUE observations. The subsequent confirmed eruption of \swuma\
began on JD 2450200, and the gap between this date and that of the 
previous confirmed outburst 
corresponds closely to twice the typical 460-day outburst cycle.   
It is conceivable that \swuma\ skipped an
outburst, but there also appear to be gaps in the coverage provided by
the AAVSO that could be large enough to miss an outburst if one had
occurred in the meantime. We can, however, state with confidence that
\swuma\ did not erupt within the timeframe extending 350 days prior to
the IUE observations. \swuma\ has also been observed in quiescence by the Hubble
Space Telescope Imaging Spectrograph (HST/STIS), an instrument with
spectral coverage similar to that of IUE SWP. The STIS data have not
yet been made public, but preview spectra are available.

\section{Discussion}\label{discuss}

When analyzing spectra in the far-UV, interstellar reddening is always
  a concern. \citet{SOH88} estimate a negligible reddening of $E(B-V)
  = 0$ for \swuma, with an upper
  limit of $E(B-V) = 0.02$ that translates into an interstellar
  hydrogen column density of $N_{H}=1\times 10^{20}$ cm$^{-3}$.
  We will therefore neglect the effects of reddening in the 
  analysis below, with the following caveat:  Even a small $E(B-V)$ can be
  significant at these wavelengths. 
In the limiting case of $E(B-V)=0.02$, the fluxes 
observed by FUSE would be
  underestimates by a factor of 1.4 \citep{CCM}. 

\subsection{Molecular Hydrogen}\label{h2}


One theory for dwarf nova outbursts is that the Magneto-Rotational
instability of \citet{BH91} 
does not operate in quiescent CV disks because of 
a very low ionization state, in which case the viscosity will be 
extremely
low.  \citet{GM98} investigated this idea, and found
that the required ionization fraction is extremely small, $\sim
10^{-6}$. Further studies have essentially confirmed this 
ionization requirement 
\citep{FSH00,KM00}, even if including the 
previously neglected effects of non-ideal MHD Hall terms 
changes the ionization threshold somewhat \citep{SS03}.
In the presence of X-rays at the observed
levels, such a low ionization state cannot be maintained
by radiative recombination in atomic gas ($\alpha\sim 10^{-13}$
cm$^3$ s$^{-1}$).
However, dissociative recombination of molecular hydrogen ions
($\alpha\sim 10^{-8}$ cm$^3$ s$^{-1}$) can provide the low
neutral fraction needed.  Thus, the theory that CV low states
result from a shutdown of the Magneto-Rotational instability
leads to an expectation that the disk of a low state CV should
be molecular.  \citet{HHS} 
have recently reported detection
of H$_{2}$ and CO infrared emission from WZ Sge in its low state.
SW UMa, like WZ Sge, has a very low accretion rate in its low state.

Pumping of H$_{2}$ by \lya\ produces H$_{2}$ emission lines in the
UV in a wide range of astrophysical systems, ranging from sunspots
\citep{J78} 
to T Tauri star accretion disks \citep{JS03,GH02} 
and Mira \citep{WKR}. 
The bands connected to the ground vibrational level fall in the 
FUSE bandpass, with the strongest features expected at about 1050 and
1100 \AA\ \citep{RBL97}.  These features are not detected in our FUSE
data, and the spectra yield
upper ($3\sigma$) limits of $3.6 \times 10^{-15}$ \cgs\ and $2.8 \times 10^{-15}$ \cgs, respectively, for their fluxes.  The \ion{C}{4} flux
observed in the archival IUE data
is $f_{\rm CIV}\approx 2\times 10^{-13}$ \cgs\ and the H$\alpha$ flux
is $f_{\rm H\alpha}\approx 1.5\times 10^{-14}$ \cgs\ (SST86). Considering
a ratio $f_{\rm Ly\alpha}/f_{\rm H\alpha}\approx 15$ or a ratio
$f_{\rm Ly\alpha}/f_{\rm CIV}\approx 2$ based on the X-ray-illuminated
accretion disk models of \citet{JCR93},
the intrinsic \lya\ flux from SW UMa before
interstellar absorption ought to be about 2--4$\times 10^{-13}$ \cgs.
The emission should arise from the disk surface, and half of it
should be directed downward if the emitting region is a simple slab.
A detailed model would be needed to predict the fraction of that
emission that excites H$_{2}$, but for a local \lya\ line width of
0.5\nolinebreak\ \AA\
(excluding Keplerian velocity shifts), approximately 10\% of the flux
is resonant with the $1-2 P(5)$ and $1-2 R(6)$ transitions of H$_{2}$, and
about 20\% of the photons absorbed should be converted to the bands
in the FUSE wavelength range (see Wood, Karovska, \& Raymond 2002).  
Thus a rough prediction for the fluxes
would be 4--8$\times 10^{-15}$ \cgs.  This is not far above the upper
limits quoted above. We also note that the 
preview STIS spectra of SW UMa do not show the H$_{\rm 2}$
bands (e.g., those near 1610 and 1670 \AA ) seen in other H$_{\rm 2}$ fluorescent
spectra.  The STIS spectra have higher signal-to-noise than the FUSE
spectra, but it is also difficult to
predict the intensity of the longer wavelength bands.
 
If the disk is indeed molecular, as indicated by the WZ Sge detections, the
lack of fluorescent emission might be attributed to a surface layer of atomic
gas
that shields the disk from \lya\ photons.  Such a layer could be a small
fraction of the disk mass, but it would have an ionization fraction much
larger than the limit given by \citet{GM98}. 
It is possible
that accretion proceeds in this surface layer, as has been proposed for
T Tauri stars by \citet{CG96} 
and developed for quiescent DNe by
\citet{KM02}. 
 
%

\subsection{Far-UV Continuum}\label{cont}

The effective temperature ($T_{\rm eff}$) of the WD in \swuma\ was
first
measured by \citet{GK99}. They fit model atmosphere spectra to
IUE SWP spectra
($\lambda = 1200$--1900 \AA) obtained during quiescence. 
Assuming a WD with surface
gravity \linebreak $\log g=8$ ($M_{\rm WD} = 0.6$ M$_{\odot}$), they derive a
best-fit effective temperature of $T_{\rm eff} = 16,000\pm 1500$\nolinebreak\ K and
estimate a distance $d=182$ pc for \swuma. Applying a similar approach
to the STIS spectra of \swuma, \citet{SSGH02} derive 
$T_{\rm eff}=14,000 K$ and $d=159$ pc. 

These values for the quiescent $T_{\rm eff}$ of the WD in \swuma\ can
be compared to that of WZ Sge, another short-period DN. 
Following an eruption in 2001 July, the WD in WZ Sge was
observed  to cool from 23,400 K
in 2001 October to 15,900 K 17 months later \citep{KL04}. 
This eruption of WZ Sge was
the first such outburst in 22 years, a timescale that is clearly much
longer than the typical outburst period of \swuma, and so it is reasonable to
expect a disparity between the cooling timescales of the two
systems. 
The WD in WZ Sge, with $M_{\rm WD} = 0.9$
M$_{\odot}$ \citep{KL03}, is more massive than the WD in
\swuma, so this may partially account for the fact that the cooling time
for WZ Sge exceeds the typical outburst period of \swuma. Given the
obvious disparities between these two CVs, it is intriguing that they
seem to
exhibit very similar $T_{\rm eff}$ in
quiescence.

Our FUSE observations of \swuma\ were taken on JD 2452220 (2001 6
November). The AAVSO lightcurves clearly indicate that an outburst
was in progress on JD 2452070, 150 days earlier. We are therefore in a
position to place a constraint on the post-outburst temperature of the WD by
investigating
the far-UV continuum emission in the region just shortward of the IUE
SWP spectral coverage. Due to the
extreme faintness of the source, the expected continuum level of the
16,000 K WD ($f_{\rm cont}\sim
5\times 10^{-15}$ \cgs\ \AA$^{-1}$ for $\lambda =$
1100--1150 \AA) of GK99 is close
to the detection threshold of FUSE, 
which is nominally 30\% of the typical detector background levels, or
$\sim 2\times 10^{-15}$ \cgs\ \AA$^{-1}$ in the LiF channels.
In order to
attempt such a sensitive measurement, we took advantage of the FUSE
``sky background'' observation (\S \ref{bkgd}). 
We performed a detailed inspection of all of the FUSE raw exposures from both
datasets 
in the wavelength regions of 1120--1130 \AA\ and
1155--1165 \AA\ of the LiF1b and LiF2a detector segments, where
the continuum is predicted to be strongest, and no strong
airglow lines are present. 
The count rates in the \swuma\ target
exposures were significantly higher, by a factor of $\sim 1.5$, than
the count rates in the sky-background exposures. This seems to suggest that
the elevated count rates are due to photons from \swuma. However,
continuum emission from a point source should produce a measureable peak in
the count density in the cross-dispersion direction of the detector,
but we do not detect any enhancement in the count density resembling a
continuum in the raw exposures. A broad component to the \ovi\
emission is
evident upon careful examination of 
the raw exposures between 1030 and 1040 \AA\ in the LiF1a
and Lif2b channels of the \swuma\ target data (\S \ref{ovi} below),
and these counts yield a peak flux of $4
\times 10^{-15}$ \cgs\ \AA$^{-1}$ in the calibrated spectra. 
The fact that we can see the broad \ovi\ in
the raw data but cannot detect a continuum at similar flux levels 
prevents us from
making a firm measurement of the far-UV continuum in \swuma. Instead,
we will place conservative upper limits on the continuum levels across
the FUSE spectral bandpass.

Working with the calibrated, fully-extracted and summed 
data from the CalFUSE pipeline, we 
subtracted the 9800-s sky-background 
spectra from the 29,000-s \swuma\ spectra.
This reduces potential 
systematic errors introduced by the FUSE background subtraction, but at
the cost of increased noise. 
Imperfections in the background subtraction
are pronounced in regions of the spectra
that lie close to a detector edge, hence we do not use these regions
when attempting to characterize the continuum. 
We further process the data by removing all prominent
airglow features and coarsely resampling the spectra in 10 \AA\ bins. We then
take
the average of the two LiF channels covering each spectral bandpass
between 1060 and 1180 \AA. We adopt twice these average values as our upper
limits on the continuum, or else we set the upper limits to $3\times
10^{-15}$ \cgs\ \AA$^{-1}$, the flux level of the significantly detected
broad \ovi\ emission, whichever is greater.
It must be stressed 
that there is no way to place a meaningful statistical upper
limit on the continuum, because the dominant sources of error are
uncertainties in the background subtraction. 
But given that the flux levels in the rebinned spectra from the
independently calibrated LiF1b and LiF2a channels
agree with each other to better than 50\%, and that we observe the
broad \ovi\ emission at a comparable flux level, we are confident
that 
our upper limits
are at least 3 times greater than the level of the systematic errors.

These upper limits, along with the IUE SWP spectrum,
are compared to a grid of WD model 
spectra provided by the authors of GK99. These models are 
based on a $\log g=8$ WD with a pure hydrogen/helium atmosphere,
whereas the models used in GK99 are for a WD with solar abundances.
This disparity causes us to adopt slightly different scale factors
when comparing the models to the data.
The scale factor used determines the ratio of
the WD radius $R_{\rm WD}$ to the distance $d$, according to the
relation
\begin{displaymath}
	f_{\rm scale}\equiv \frac{f}{F}=\frac{\pi R_{\rm
WD}^{2}}{(3.086\times 10^{18} d)^{2}},
\end{displaymath}
where $f$ is the observed flux and $F$ is the model flux.
We test models for $T_{\rm eff}$ in the range of 12,000--18,000 K (in
increments of 1,000 K).
We calculate $f_{\rm scale}$ by matching the model flux to the IUE
flux within the region of 1250--1500 \AA,
accepting as
plausible  values of $f_{\rm scale}$ that give 84 pc $\le d \le$ 243
pc for a $\log g=8$ WD.

\begin{figure}
\epsscale{.80}
\plotone{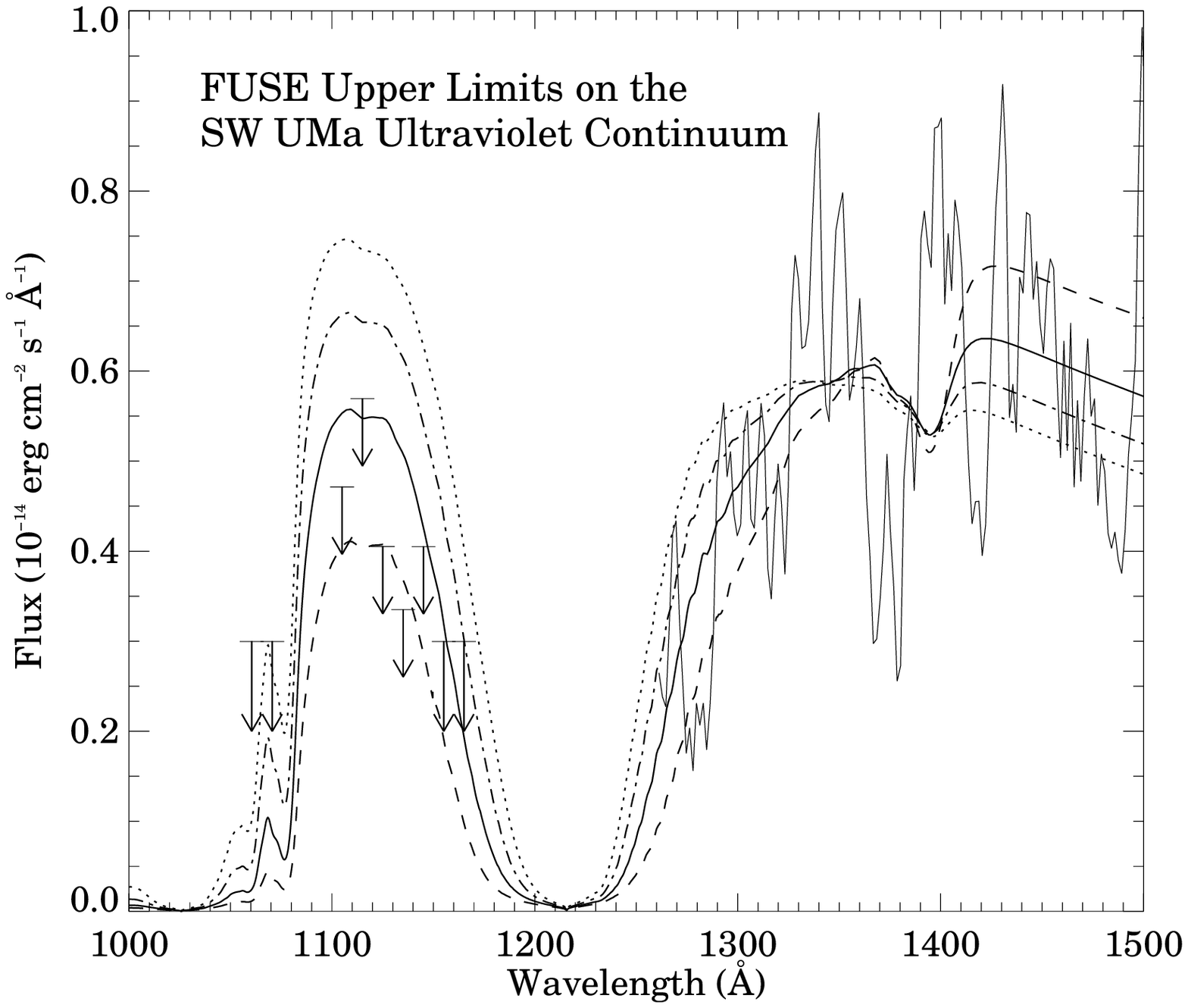}
\caption{Effective temperature of the white dwarf from the UV
spectrum. 
The FUSE upper limits on the far-UV continuum in the range of
1060--1180 \AA\
are shown by arrows; these, along with
the IUE SWP spectrum of \swuma\ for 1250--1500 \AA, are compared to
model spectra scaled to match the integrated flux from IUE. The models
are
calculated
for a pure hydrogen--helium WD, and their most striking features are the
extremely broad \lya\ absorption at 1216 \AA\ and sharp drop in flux
shortwards of 1100 \AA. 
The model that best agrees with both the FUSE upper limits and the IUE data
 is for $T_{\rm eff}= 15,000$ K (solid line). Also shown are models for
14,000 K (dashed line), 16,000 K (dash-dotted line), and 17,000 K (dotted line).
\label{contfig}}
\end{figure}

The results of this analysis of the UV continuum of \swuma\ are shown
in Fig. \ref{contfig}. We 
find that a model with $T_{\rm eff}\le 15,000$ K for a 0.6
M$_{\odot}$ WD at $d=173$ pc best matches the IUE spectrum while
falling within the upper limits on the far-UV continuum imposed by FUSE.
Hence, the effective temperature
of the WD 
does not exceed 15,000 K at the time of
the FUSE observations, 150 days post-outburst. 
This result agrees well with the findings of SSGH02 and 
with the lower end of the temperature range
derived 
by GK99 from the IUE spectrum, 
350 days post-outburst (\S
\ref{AAVSO} above), but the flux in their 16,000 K model significantly
exceeds our
conservative upper limits in the far-UV.


\subsection{On the Origin of the \ovi\ Emission Lines}\label{ovi}

\begin{figure}
\epsscale{.90}
\plotone{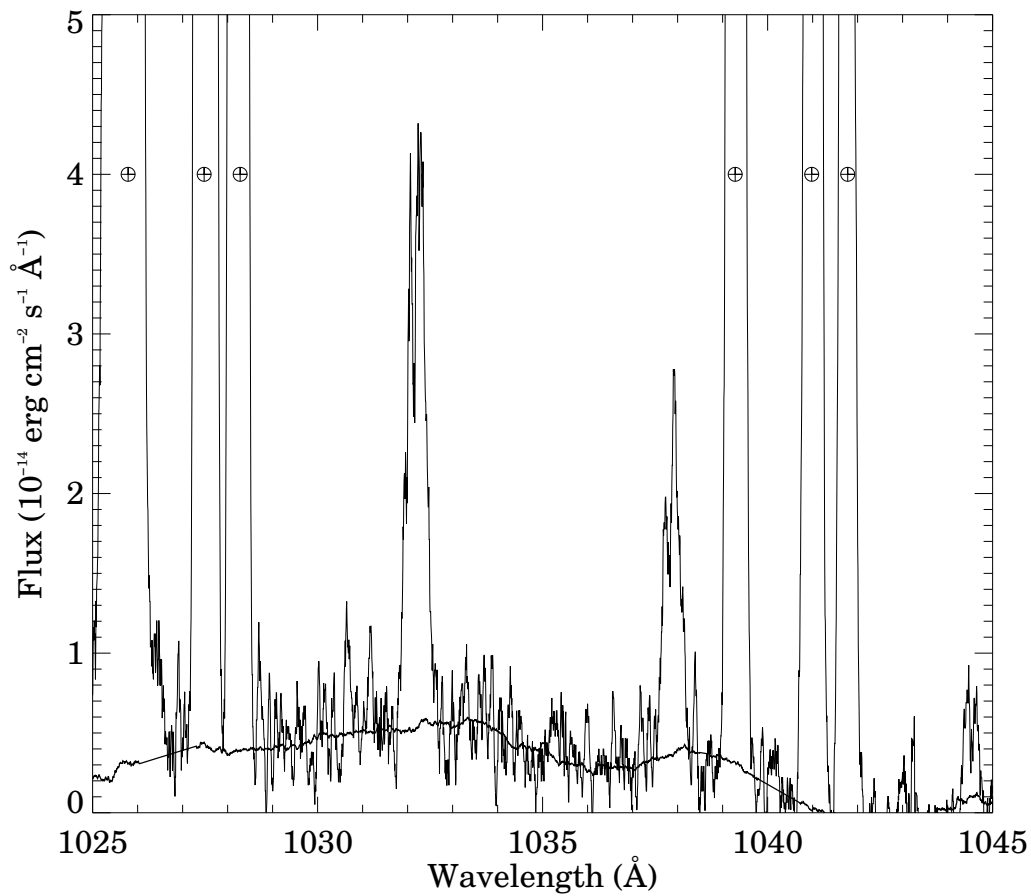}
\caption{Region 
of the FUSE LiF1a channel containing the \ovi\ 1031.9 
and 1037.6 \AA\ emission features. The narrow emission lines
exhibit redshifted, asymmetric, double-peaked profiles. The \ovi\
region is surrounded by strong airglow lines of \ion{O}{1} and
Ly$\beta$. 
A heavily
smoothed spectrum with the airglow lines and the narrow \ovi\ removed
has been overplotted (heavy line) to highlight the underlying broad
component of the \ovi\ 1031.9 emission; a broad component of \ovi\
1037.6 is also 
suggested but is excluded from the analysis because of its extreme
faintness and proximity to the \ion{O}{1} 1039.2 \AA\ airglow line.
\label{ovilines}}
\end{figure}

The narrow \ovi\ lines at 1031.91 and 1037.61 \AA, with integrated 
fluxes of $1.8\times
10^{-14}$ \cgs\ and $8.9\times 10^{-15}$ erg
cm$^{-2}$ s$^{-1}$, respectively, are the most strongly detected
emission lines in the FUSE spectra of \swuma\
(Fig. \ref{ovilines}).  
These lines are nevertheless far too faint to allow
phase-sampling of the 81.2-minute binary orbital 
period with only 29,000 s of FUSE
exposure time. Thus, the observed profiles
represent an integration over multiple orbits. As such, these lines
are puzzling in several respects. They are very narrow, with FWHM of 150
\kms. 
They exhibit
unusual, double-peaked shapes that are slightly
asymmetric. And they appear to be redshifted by $\sim 80$ \kms\ with
respect to heliocentric rest. 
In order to investigate the possible sources of the \ovi\ emission
in \swuma, we consider the plausibility of an origin in each of the
three main components of this binary system; the primary, the
secondary, and the disk/accretion stream.

{\it The secondary.}  The companion star in the \swuma\ system is
believed to be a very low-mass, late-type (0.1
M$_{\odot}$, $B-V=1.7$) main-sequence star (SST86). It 
is almost certainly
too faint in the UV to be observable by FUSE. To quantify this, we assume
that the companion has $L_{\rm bol} \sim 10^{-3}$ L$_{\odot}$ and is
at a
distance of $d=182$ pc. The observed bolometric luminosity at Earth will
thus be 
$l_{\rm bol}=9.64 \times 10^{-13}$ erg cm$^{-2}$ s$^{-1}$. Given that
the luminosity in \ovi\ should be only a tiny fraction ($\sim
10^{-5}$) 
of the bolometric luminosity, it is extremely unlikely that the
secondary contributes to the observed \ovi\ lines. Furthermore, the
most plausible orbital parameters for the system, when integrated over
the orbital period of the system, would yield spectral
lines approximately twice as wide as those observed here.

{\it The disk/accretion stream.} Based on observations of other CVs, it 
is entirely reasonable to expect that \ovi\
emission could originate in the disk, a hotspot in the disk, or even
in the accretion stream flowing from the secondary. Disk emission almost
certainly dominates the \ovi\ line profiles during outburst, as seen
in WZ Sge \citep{KL03}, but to date there have been no FUSE
observations of \swuma\ during an outburst phase.  A very broad \ion{C}{4}
1550 \AA\ emission feature dominates the archival 
IUE SWP quiescent spectra of
\swuma, with a flux $f_{\rm CIV}\approx 2\times 10^{-13}$ \cgs\ and a
 line width of $\sim 2000$ \kms\ that is comparable to
that of the H$\alpha$ emission line observed by SST86. The H$\alpha$
profiles show a doubled-peaked shape characteristic of an origin in a
rotating Keplerian 
disk, but this level of detail in the \ion{C}{4} profile is
not resolved by IUE (in any case, the \ion{C}{4} feature is a
blend). In contrast, the narrow \ovi\ lines observed by FUSE, although
exhibiting a double-peaked shape, are only 150 \kms\ wide, a value
that is an order of
magnitude smaller than  
the typical Keplerian orbital velocities in the disk.
As noted in \S \ref{cont}, the data do show a faint, broad
component to the \ovi\
emission at 1032 \AA, $\sim 2000$ \kms\ wide, with an integrated flux of
$f_{\rm broad}=2\times 10^{-14}$ \cgs\ $\pm 40\%$. 
The large error on this
value is primarily due to the presence of the strong airglow lines of
\ion{O}{1} and Ly$\beta$ that bracket the broad
\ovi\ emission (see Fig. \ref{ovilines}). 
While most of the airglow emission can be removed from
the spectrum, residual contamination from the line wings still prevents 
an accurate measurement of the true 
zero-flux level on either side of the broad \ovi.
Nevertheless, the ratio of the broad \ovi\ to \ion{C}{4} emission is
small, with
$f_{\rm broad}/f_{\rm CIV}\sim 1/10$. The \ciii\ emission at 1176
\AA\ is similarly broad and comparably bright, with an integrated flux
of $f_{\rm CIII}=2.5\times 10^{-14}$ \cgs\ $\pm 20\%$.
This suggests that the \ion{C}{4} and \ciii\
emission from the disk originates in relatively cool, quite likely
photoionized gas \citep{YK96,MLK97}. 

We have run a similar photoionization model \citep{JCR93} 
using the X-ray luminosity of SW UMa for an annulus near
the outer edge of the disk ($r=10^{10}$ cm).  It predicts 
$f_{\rm broad}/f_{\rm CIV}=1/40$, in keeping with the small
value observed.  The \ciii\ 977 \AA\ intensity is
predicted to be 8 times the \ovi\ intensity, and
\ciii\ 1176 is predicted to be twice as
bright as \ovi.  While these numbers are consistent
with the observed fluxes only at the factor of 2 level,
the model code, which was designed for low mass X-ray
binaries, is pushing the range of its validity, and
the model parameters are uncertain.  No general model
of UV emission from a  magnetically heated disc
atmosphere is available, but the solar transition
region shows an \ovi\ to \ion{C}{4} ratio of about 1/3 to 1/2,
while the ratios of \ovi\ to \ciii\ 977 \AA\ and
\ciii\ 1176 \AA\ are about 2/3 and 3/2, respectively \citep{WR02}.

The possibility remains that the narrow \ovi\ emission comes from a
localized area in the disk or accretion stream. 
The most likely source of such emission would be the hot spot
where the accretion stream joins the outer edge of the disk. 
We have modeled the line profile by assuming
that the emitting source orbits the center-of-mass of the binary
system at a velocity typical of the outer edge of the accretion disk,
and that the emission is anisotropic, visible to FUSE only during the
redshifted phase of the orbit. This can reproduce the 
observe redshift and line width, but not the double-peaked shape.
The anisotropy could potentially arise if the hot
spot were obscured by the disk or accretion stream at viewing angles
corresponding to the blueshifted orbital phase. There are
problems with this interpretation, however. 
The 
condition that the emission be visible only during certain convenient
orbital phases is arbitrary, and the orbital velocity chosen is poorly
constrained. Worse, the observed brightness of the
narrow \ovi\ lines is very difficult to explain in terms of a hotspot in a
quiescent DN. The interaction of the accretion stream with the disk
edge is highly unlikely to produce a shock strong enough to emit
\ovi\ at the observed levels. 
An alternative is that the \ovi\ comes from photoionized
gas. But because the narrow and broad components of the \ovi\ emission
exhibit similar total fluxes, photoionization
would require that the hotspot or small region of the accretion stream
emitting the narrow \ovi\
component present a solid angle to the WD comparable to that of the
entire accretion disk. This condition seems implausible at best. 
We thus conclude that while the broad \ovi\ emission comes from the disk,
the narrow \ovi\ emission probably originates elsewhere. 

{\it The primary.} 
The remaining plausible source of the \ion{O}{6} emission is
the white dwarf itself.  
Chandra X-ray spectra of non-magnetic CVs with low accretion
rates and also of the intermediate polar WX Hya are matched reasonably well  
by models of gas cooling from a temperature around 20 keV \citep{PS02,KM03,RP03,H04}.
Barring an
implausible termination of the cooling at a temperature above $10^{6}$ K,
such cooling flows must also produce \ion{O}{6}. 
The cooling flow is essentially
identical to the cooling behind a shock front, so an extended version of
the shock models of \citet{JCR} 
can be used to predict the X-ray and UV
emission spectrum. A model for a 4250 \kms\ shock (peak temperature
$T=20$ keV)
with a preshock density of $n=10^{10}$ cm$^{-3}$, matches the X-ray
spectrum of V426 Oph \citep{H04} 
and predicts an \ovi\ flux of
$2.3 \times 10^{8}$ \cgs\ out the front of the shock.
For a WD of radius $R_{\rm WD}=8.9\times 10^{8}$ \nolinebreak cm at a distance of 182 pc (GK99), 
this implies a flux at the
Earth of $5.7 \times 10^{-16}$ \cgs\ 
and an accretion rate of $7.0\times 10^{13}$
g s$^{-1}$.
The \ovi\ flux is directly proportional to the preshock density and thus to
the accretion rate. The observed flux of $2.7\times 10^{-14}$ 
\cgs\ in the narrow \ovi\ component indicates an accretion 
rate of $3.0\times 10^{15}$ g s$^{-1}$. 

The shock model also predicts
that the ratio of the X-ray to the \ovi\ luminosity will be
$L_{\rm X}/L_{\rm OVI}
= 1.7\times 10^{3}$ for $T=20$ keV. SST86 use their EXOSAT Channel
Multiplier Array 
observations of \swuma\ to derive soft (0.05--2 keV) X-ray fluxes that range from
$3.2\times 10^{-12}$ \cgs\ for a 0.07 K blackbody model to 
$5\times 10^{-12}$ \cgs\ for a 0.21 keV thermal bremsstrahlung
source. \swuma\ was also marginally ($2\sigma$) detected by the the
EXOSAT medium energy detector in the range of 2--6 keV, with a limit of 
$7\times 10^{-12}$ \cgs\ placed on the flux for a 10 keV thermal
bremsstrahlung spectrum. Taken together, the EXOSAT X-ray fluxes, when
compared to the FUSE \ovi\ fluxes, suggest that $L_{\rm X}/L_{\rm
OVI}\sim 300$. If instead we assume a 20 keV bremsstrahlung, as
in the V426 Oph model, the X-ray flux is doubled while the flux in
\ovi\ remains constant, and the ratio of luminosities becomes 
$L_{\rm X}/L_{\rm OVI}\sim 600$. 
\citet{R94} provide a further check on the soft
X-ray luminosity of \swuma. They derive a flux of $2.7\times 10^{-12}$ \cgs\
in the 0.1--2.5 keV range by fitting a two-component \citet{RS77} 
model to a ROSAT spectrum obtained over a period of 25 days
beginning 4 days after the cessation of
an outburst. 
Given the inconsistencies between the various
models fit to the EXOSAT data by SST86, the unknown 
attenuation of the observed X-rays by interstellar material, and the
indisputable variability of the source, the error on these estimates
is undoubtedly large, a factor of 2 or more. 
Within the boundaries of these uncertainties, however,
the \ovi\ emission observed by FUSE appears to be 
consistent with a cooling flow
onto the surface of the WD.

One caveat to this interpretation must be
mentioned. SSGH02 give a rotational velocity of $v\sin i=200\pm 50$
\kms\ for the WD in \swuma. This value is based on a
model atmosphere spectrum that fits several metallic absorption
lines in the STIS spectra of \swuma. The value of $200$ \kms\ is not
very different from the widths of the narrow \ovi\ components, but if
of the intrinsic \ovi\
emission line is actually rotationally broadened by $200$ \kms, 
then the emission profile observed by FUSE, as the integration over many orbits,
cannot present a two-peaked
shape, as explained in \S\ref{details} below. Hence, if both the high
rotational velocity for the WD and the double-peaked line shape are
correct, then we would expect that the accretion from the cooling flow
might not be uniformly 
distributed over the surface of the WD, but could instead be
concentrated at the magnetic poles at intermediate latitudes.

\subsection{Details of the \ovi\ Emission Line Profiles}\label{details}

The redshifted line centroid and the double-peaked, asymmetric 
line profiles remain difficult to explain.  In the shock
model, the gas has slowed to a few \kms\ by the time it
reaches the temperature range of $10^{5}$ to $10^{6}$ K where \ovi\ is
formed.  However, the cooling flow in the shocked gas becomes
violently thermally unstable in this temperature range. This can
lead to larger downflow speeds due to lack of pressure support and
to random motions at the level of tens of \kms\ ({\it e.g.}
Innes 1992).  This may or may not be able to produce the $\sim 80$
\kms\ redshift observed in the \ovi\ lines. The instability also affects the 
predicted emission line fluxes, but probably only by about a factor of
2 when averaged over space and time. 
In any case, it is unclear how the heliocentric 
redshift of $\sim 80$ \kms\ 
 translates
into the rest frame of 
the center-of-mass of the \swuma\ system, because the
proper motion of the system is not very well 
constrained.\footnote{The currently accepted
value for the heliocentric radial velocity of \swuma\ is -20
\kms. This value appears to have been taken from the systemic velocity parameter
used in fits to the wings of the broad 
H$\alpha$ line profiles performed by
SST86.} It may be that the redshifts of the narrow \ovi\ lines
themselves are
indicative of a high proper motion for the system, but we cannot make
this claim 
solely on the basis of two relatively faint lines from a single species of ion
in the FUSE spectra.

We have further investigated the double-peaked structure of the \ovi\
lines by refining the CalFUSE extraction procedure. 
Both \ovi\ lines are observed over 
small regions on the LiF1a detector with enhanced 
number density of pixel counts 
compared to scattered light (or ``background'' counts). 
The number density of \ovi\ counts over 
these regions exhibits a significant peak in the cross-dispersion
direction. We can therefore decrease the height of the spectral box used 
by CalFUSE v2.2.3 to extract the 1-D spectrum. 
This limits small background contributions 
in the cross-dispersion direction and further increases 
the signal-to-noise (S/N) 
around the \ovi\ emission lines. The procedure is equivalent to limiting 
the height of the LWRS aperture to screen out scattered light contributions
from areas against the sky that are not filled by the point source as its
spectrum is focused on the detector. 
%
To determine a mean noise 
level in the emission lines, we construct a histogram of the difference between 
the extracted spectrum and 
the same spectrum smoothed by convolution with a gaussian of
0.09 \AA\ HWHM. The resulting flux distribution provides 
a 1-$\sigma$ (rms) noise value of $7.6\times 10^{-15}$ \cgs\ \AA$^{-1}$ 
in LiF1a. Since we 
observe that the central depths in the smoothed spectrum of the \ovi\ 
lines exceed $1\times 10^{-14}$ \cgs\ \AA$^{-1}$
with respect to their flux maxima, we conclude 
that these central depressions are statistically significant. 
The rms value of the spectral noise is also smaller than the 
difference between the smoothed peak flux in the red and violet emission 
component of \ovi\ 1031.9 \AA, signaling an asymmetric line profile. 
Our advanced recalibration procedure of the LiF1a 
channel yields somewhat smaller noise levels for the central depression of 
the \ovi\ 1031.9 \AA\ emission line, thereby confirming the double-peaked,
asymmetric shape of the line profile. 

\begin{figure}
\epsscale{.90}
\plotone{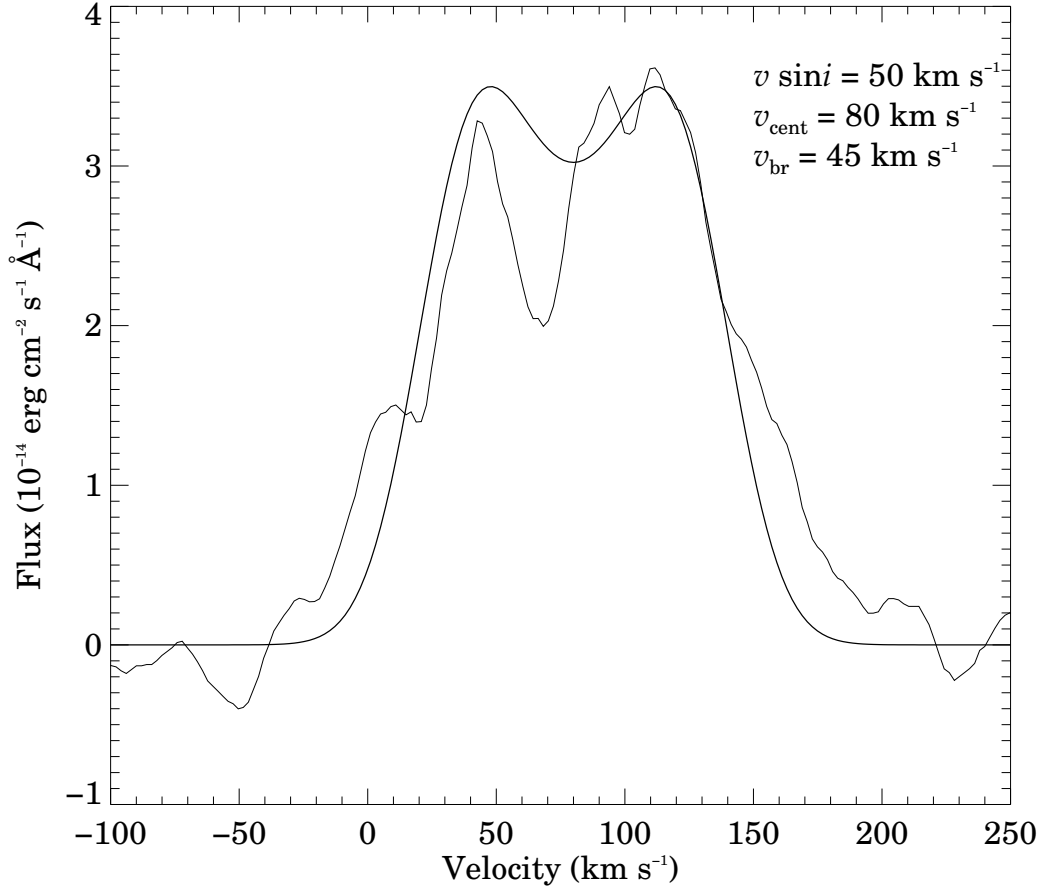}
\caption{Comparison of a simulated  
line profile based on orbital
geometry and dynamics with the narrow component of
the \ovi\ 1031.9 \AA\ emission from the
FUSE LiF1a data. The broad \ovi\ component has been
subtracted from the spectrum. 
We assume that 
a gaussian emission line of width $v_{\rm br}=45$ \kms\ (a value which
incorporates plausible values for thermal and rotational
broadening) originates from a WD in a binary orbit with radial velocity
semiamplitude $v\sin i=50$ \kms. If the line is integrated through at least
one orbital period, the result is a double-peaked profile. The
simulated profile is scaled to match the observed flux and
given a redshift $v_{\rm cent}=80$ \kms,
the velocity of the observed displacement of
the line centroid with respect to 
heliocentric rest.
\label{dynamodel}}
\end{figure}

A double-peaked shape is a signature of emission
lines formed in a
rotating disk or annulus. In such cases, 
the separation of the two peaks  
corresponds to the Keplerian velocity at the outer edge of the disk,
while the overall width and shape of the profile are governed by the velocity
of the inner edge of the disk and the thermal/turbulent velocities
within the emitting material. Although the \ovi\ lines in \swuma\ are
too narrow, with the peaks too close together, to have their origin in
the disk, in general an orbiting/revolving emitting region can
produce a similar profile, provided that the integration time spent
observing the line exceeds the period of revolution, 
and that the thermal/turbulent velocities present in the
emitting material do not exceed the bulk velocity
due to the revolution. We have simulated an emission line profile
produced by a gaussian emitter moving in a circular orbit with a
radial velocity semiamplitude of $v\sin i=50$ \kms, 
the value derived by SST86
for the WD in \swuma. This model is symmetric, so it cannot reproduce
the asymmetry observed in the profiles, and the redshift ($v_{\rm
cent}=80$ \kms) is forced to
match the centroid of the observed lines. The model is scaled so that
the simulated line profile has the same integrated 
flux as the observed line profile, as shown in Fig. \ref{dynamodel}.
We find that this approach reproduces the
overall width of the line and the separation of the peaks 
reasonably well, but
overestimates the flux in the line center while
underestimating the flux in the line wings.

\section{Conclusions}

\swuma\ has presented a challenge for observers and theorists of
DNe alike. It is a faint target and seems to exhibit some 
unusual characteristics that complicate its classification within the
standard CV taxonomy. The observations and analyses presented here
represent a first look at \swuma\ in the far-UV wavebands covered by
FUSE. The most interesting features in the FUSE spectra are the
emission lines of \ovi\ at 1031.9 and 1037.6 \AA. These lines possess both narrow
and broad components. 

The broad \ovi\ and \ciii\ 1176 \AA\ emission suggest a thin, photoionized layer
on the surface of the accretion disk in \swuma, which is exposed to
X-ray radiation from the WD. The upper limits
on the H$_{2}$ emission are consistent with a layer of atomic H immediately
below the \ovi\ emitting region that can shield the bulk of the disk,
which could consist of cooler, molecular gas. Indeed, the presence of
atomic H is a necessary condition to enable the mechanism of layered
accretion in DNe (see Menou  2002). 
Given that H$_{2}$ has been detected in IR spectroscopy of WZ Sge
\citep{HHS}, a search for H$_{2}$ emission from \swuma\ in the IR
could explore the disk structure in more detail.

The measured fluxes of the narrow \ovi\ emission
profiles
are consistent with
formation in a cooling flow onto the surface of the WD and suggest a
quiescent accretion rate for the system of $\sim 3\times 10^{15}$
g s$^{-1}$. The double-peaked nature of the line profiles can be
accounted for, at least partially, by the orbital motion and geometry
of the WD, but the
asymmetries and redshifts in the line profiles remain difficult to
explain. We cannot rule out an origin of the narrow \ovi\ in either the hot
spot or some other localized area
of the disk or accretion stream, but find this interpretation less
compelling than that of a cooling flow onto the WD. Either phase-resolved
far-UV spectroscopy of \swuma\ or a reliable determination
of the systemic
velocity could help settle these
outstanding questions.

It is also a bit puzzling that we do not detect the 
\ion{C}{3} emission line at 977 \AA, given  
the presence of 
the \ciii\ emission around 1176 \AA. We searched for evidence of this
line in the spectra from the SiC1b and SiC2a channels of FUSE, but
found no significant emission at 977 \AA\ in SiC2a and 
only a weak profile in SiC1b. The latter is due to a known
contamination from scattered solar \ion{C}{3} 977 \AA\ in that channel. The SiC
channels of FUSE are generally more vulnerable to scattered light
contamination than are the LiF channels, and the
sensitivity in the region of 977 \AA, covered only by the SiC
channels, 
is down by a factor of 2 from
the sensitivity of the LiF channels in the vicinity of 1032 \AA. Hence,
the non-detection
emission from \ion{C}{3} 977 \AA\ and other, weaker resonance lines in the
FUSE spectra of \swuma\ is a function of the extreme 
faintness of the source coupled with the variation in detector sensitivity with
spectral channel, and does not exclude the possibility of an
abundance 
of other 
ionic species in the \swuma\ system.

The upper limits on the faint, far-UV continuum limit the
temperature of the WD during the quiescent phase to less than 15,000
K. This result agrees well with the $T_{\rm eff}=14,000$ K of SSGH02
and places a downward constraint upon 
the estimate of $T_{\rm eff}=16,000 \pm
1500$ K derived by GK99 from the IUE spectrum of \swuma. 

It might be worthwhile to attempt a
target-of-opportunity FUSE observation of \swuma\ during or
immediately following a DN eruption
of the system. 
The
high-state temperature of the system should yield a
far-UV continuum level that would be easily observable by FUSE,
allowing a measurement of the post-outburst cooling in \swuma\ similar
to that already carried out for WZ Sge \citep{KL03}.
In addition, such observations 
would enable a comparison of the \ovi\ profiles during
outburst with the quiescent profiles presented here, potentially
leading to a more complete
description of the physical mechanisms driving the 
accretion in this system
over the course of the outburst cycle.






\acknowledgments

We are grateful to Detlev Koester for providing us with model white
dwarf thermal spectra. We are indebted to the late Janet Mattei
and the AAVSO for their vigilant monitoring of \swuma. This work was
supported by NASA grant NAG5-10353. AL thanks P. Young at RAL (UK)
for helpful discussions on FUSE calibration procedures. We thank the
anonymous referee for a thorough review and insightful comments. FUSE is
operated for NASA by Johns Hopkins University under NASA contract NAS5-3298.


\clearpage



\end{document}